\documentclass[12pt]{article}
\usepackage[scale=0.85]{geometry}
\usepackage{graphicx,amssymb,amsmath}
\usepackage[english]{babel}
\usepackage{authblk}
\usepackage{cite} 
\usepackage{setspace}

\let\jnlstyle=\slshape
\def\refjnl#1{{\jnlstyle#1}}%
\newcommand\aj{\refjnl{Astron.~J.}}%
\newcommand\araa{\refjnl{Annu.~Rev.~Astron.~Astrophys.}}%
\newcommand\apj{\refjnl{Astrophys.~J.}}%
\newcommand\apjl{\refjnl{Astrophys.~J.~Lett.}}%
\newcommand\apjs{\refjnl{Astrophys.~J.~Suppl.~Ser.}}%
\newcommand\aap{\refjnl{Astron.~Astrophys.}}%
\newcommand\mnras{\refjnl{Mon.~Not.~R.~Astron.~Soc.}}%
\newcommand\pasp{\refjnl{Publ.~Astron.~Soc.~Pacif.}}%
\newcommand\nat{\refjnl{Nature}}%
\newcommand\jcp{\refjnl{J.~Comput.~Phys.}}%

\newcommand\kmps{\mbox{km\,s${}^{-1}$}}%
\newcommand\Msun{\mbox{$M_\odot$}}%
\newcommand\arcdeg{\mbox{$^\circ$}}%
\newcommand\arcsec{\mbox{$^{\prime\prime}$}}%
\newcommand\farcs{\mbox{$.\!\!\arcsec$}}%

\title{The large-scale nebular pattern of a superwind binary in an eccentric orbit}
\author[1,5]{Hyosun Kim*}
\author[1]{Alfonso Trejo}
\author[1]{Sheng-Yuan Liu}
\author[2]{Raghvendra Sahai}
\author[1,3]{Ronald E.\ Taam}
\author[4]{Mark R.\ Morris}
\author[1]{Naomi Hirano}
\author[1]{I-Ta Hsieh}

\affil[1]{Academia Sinica Institute of Astronomy and Astrophysics, 
  P.O. Box 23-141, Taipei 10617, Taiwan}
\affil[2]{Jet Propulsion Laboratory, California Institute of Technology
  MS 183-900, 4800 Oak Grove Drive Pasadena, Ca 91011, USA}
\affil[3]{Department of Physics and Astronomy, Northwestern University, 
  2145 Sheridan Road, Evanston, IL 60208, USA}
\affil[4]{University of California, Los Angeles, CA 90095-1547, USA}
\affil[5]{EACOA fellow}
\date{*Corresponding author --- Hyosun Kim (hkim@asiaa.sinica.edu.tw)}

\begin{document}
{\begin{center}\bf\Huge The large-scale nebular pattern of a 
superwind binary in an eccentric orbit\\[20pt]\end{center}}

{\setlength{\parindent}{0in}

{\bf\large Hyosun Kim$^{1,5,*}$, Alfonso Trejo$^1$, Sheng-Yuan Liu$^1$, 
Raghvendra Sahai$^2$, Ronald E.\ Taam$^{1,3}$, Mark R.\ Morris$^4$,
Naomi Hirano$^1$ \& I-Ta Hsieh$^1$\\[10pt]}
$^1$ Academia Sinica Institute of Astronomy and Astrophysics, 
  P.O. Box 23-141, Taipei 10617, Taiwan\\
$^2$ Jet Propulsion Laboratory, California Institute of Technology
  MS 183-900, 4800 Oak Grove Drive Pasadena, Ca 91011, USA\\
$^3$ Department of Physics and Astronomy, Northwestern University, 
  2145 Sheridan Road, Evanston, IL 60208, USA\\
$^4$ University of California, Los Angeles, CA 90095-1547, USA\\
$^5$ EACOA fellow\\
* Corresponding author --- Hyosun Kim (hkim@asiaa.sinica.edu.tw)
}\\

{\setlength{\parindent}{0in}
\textbf{%
Preplanetary nebulae (pPNe) and planetary nebulae (PNe) are evolved, 
mass-losing stellar objects that show a wide variety of morphologies. 
Many of these nebulae consist of outer structures that are nearly 
spherical (spiral/shell/arc/halo) and inner structures that are highly 
asymmetric (bipolar/multipolar) \cite{bal02,sah11}. 
The coexistence of such geometrically distinct structures is enigmatic 
because it hints at the simultaneous presence of both \emph{wide} and 
\emph{close} binary interactions, a phenomenon that has been attributed 
to stellar binary systems with eccentric orbits \cite{kim15}. 
Here we report new high-resolution molecular-line observations of the 
circumstellar spiral-shell pattern of AFGL 3068, an asymptotic giant 
branch (AGB) star transitioning to the pPN phase. The observations 
clearly reveal that the dynamics of the mass loss is influenced by the 
presence of an eccentric-orbit binary. 
This quintessential object opens a new window on the nature of deeply 
embedded binary stars through the circumstellar spiral-shell patterns 
that reside at distances of several thousand Astronomical Units (AU) 
from the stars.}\\
}

\renewcommand{\figurename}{{\bf Figure}}
AFGL 3068, an extreme carbon star at the tip of the AGB evolutionary 
phase, is a remarkable source with the best-characterized, complete 
spiral pattern in its circumstellar envelope (CSE). This unambiguous 
spiral pattern was the first ever revealed surrounding an evolved star 
in a dust-scattered light image in the optical band (at $0.6\,\mu$m) 
of the Hubble Space Telescope (HST) \cite{mor06,mau06}. 
The striking discovery of the presence of this very well-defined pattern 
has prompted new research on how binarity can affect mass outflows during 
late stages of stellar evolution (AGB, pPN, and PN). In particular, recent 
theoretical investigations have shown that such patterns can naturally be 
explained by the orbital motion of a mass-losing star in a binary system 
\cite{sok94,mas99,kim12a,kim12b,kim13}. 
In the case of AFGL 3068, there are indeed two point-like sources in its 
central region detected with Keck adaptive optics near-infrared imaging, 
revealing a projected binary separation of 109\,AU \cite{mor06}. 
Constraints on its binary parameters have been derived on the basis of 
these HST and Keck images, assuming a circular orbit \cite{kim12b}. 
This further indicated that the degeneracy imposed by the two-dimensional 
image of the three-dimensional structure can be lifted by high-resolution 
molecular line observations.

Our new observations of AFGL 3068 taken with the Atacama Large 
Millimeter/submillimeter Array (ALMA; see Methods for details on 
observations and data calibrations) unveil exceptionally detailed 
features in its CSE (Fig.\,\ref{fig:chm}; individual molecular lines 
are presented in Supplementary Figs \ref{fig:12CO}--\ref{fig:HC3N}).
A spiral pattern is definitively detected over a radius ($r$) of 10\arcsec, 
corresponding to 10,000\,AU at the distance of AFGL 3068 ($\sim$\,3,400 
light years) in the $^{12}$CO $J=2-1$ and $^{13}$CO $J=2-1$ molecular lines 
(see the middle panel of Fig.\,\ref{fig:chm}). The HC$_3$N $J=24-23$ line 
best highlights the innermost winding of the spiral pattern. The emission 
maps integrated over the molecular lines are well correlated with the 
HST image, thus verifying that the circumstellar dust and molecular 
gas trace the same spiral feature. Remarkably, the molecular line maps 
reveal the presence of the innermost winding of the spiral ($r<3\arcsec$), 
which was absent in the dust-scattered light image. 

The observed emission pattern follows approximately a straight line when 
displayed in the radius ($r$) versus angle ($\phi$) plot 
(Fig.\,\ref{fig:trd}). 
Such a projected shape is markedly similar to an Archimedean spiral to 
first-order. 
Hydrodynamic models show that a perfect Archimedean spiral pattern forms 
in the CSE surrounding a mass-losing star in a circular orbit, viewed
with the orbital plane located near the plane of the sky \cite{kim12a}. 

While the molecular line emission near the systemic velocity exhibits 
a remarkable spiral pattern (Fig.\,\ref{fig:chm}, middle row panels), 
the emission in channels near the expansion velocity shows rather 
ring-like patterns (top and bottom panels). 
This spiral-ring bimodality has been predicted as the characteristic 
of binary-induced spiral-shell models having their orbits tilted with 
respect to the sky plane \cite{kim13}. 
The spiral-shell models further suggest that ring-like patterns found 
in optical images of many AGBs, pPNe, and PNe possibly originated from 
binary systems, but viewed at a range of inclination angles 
\cite{mas99,kim12a,kim12b,kim13}. 

The most exceptional feature of the images presented here is the 
previously unrecognized bifurcation in the spiral pattern (i.e., a 
separation and divergence of the otherwise single-stranded pattern 
at a given position angle). 
A bifurcation is visible at the second winding ($r\sim4\arcsec$ to 
the North-East in the central velocity channel) and shifts toward 
smaller radii at larger velocity channels. 
A similar bifurcation structure is also found at the first winding of 
the spiral at $r\sim2\arcsec$ (see Fig.\,\ref{fig:cmp}a, dotted lines). 
These bifurcation features appear prominently in the angle--radius plot 
(Fig.\,\ref{fig:cmp}b) as revealed by the descending trend, superimposed 
on the ascending trend of the main spiral pattern. 

We suggest that a binary system in an eccentric orbit provides the most 
plausible explanation for the bifurcation seen in AFGL 3068. Following 
Kepler's second law, the speed of the star is fastest at the periastron, 
thus boosting the wind material most strongly in the direction of motion. 
The resulting wind gusts produced at periastron overtake the material
expanding with slower speeds that was ejected throughout the orbit by 
spherical mass loss from the AGB star (see Supplementary Fig.\,\ref{fig:pw}). 
This leads to a spatial distribution of gas that is manifested as a local 
bifurcation. 
Fig.\,\ref{fig:cmp} presents a comparison between the molecular emission 
at the systemic velocity observed by ALMA (top) and the corresponding gas 
mass distribution within the same velocity range in a hydrodynamic model 
(bottom; see Methods for the hydrodynamic model setup). 
The model of a binary system in an eccentric orbit generates bifurcations 
that are coincident with those present in the observed images.

Bifurcations of the pattern have never appeared in previous binary models 
based on circular orbits.
In a circular orbit model with the orbital plane inclined with respect 
to the plane of the sky, the spiral-shell can exhibit an undulation 
(with respect to the straight line characterizing an Archimedean spiral) 
in the angle--radius plot \cite{kim13}. 
However, there is no inclination angle at which a bifurcation is produced. 

Three-dimensional hydrodynamic simulations, assuming zero orbital 
eccentricity, showed that the width of the spiral pattern increases 
with radial distance \cite{kim12a}. As a result, the inner and outer 
edges of the thickened spiral pattern may resemble a bifurcation. 
The growth rate of the pattern width, however, is proportional to the 
ratio of the local sound speed to the wind expansion speed, which is 
typically less than 10\% over the CSEs of AGB stars. Therefore, the 
separation between the inner and outer edges of the pattern would be 
unresolvable, as 10\% of the arm-to-arm interval in AFGL 3068 is less 
than 0\farcs3. 
More critically, both edges of the spiral arm should be ascending in the 
angle--radius plot, unlike the descending trend seen for the bifurcated 
features of AFGL 3068. 

In another study, a periodic mass loss from a star was introduced 
to mimic the effect of episodic close interactions with its binary 
companion \cite{cer15}. This model with the modulated gas densities, 
and hence the modulated molecular line intensities, produced very 
complex overlapping ring patterns rather than well-defined bifurcation 
features in the CSE. The dynamical influence induced by the non-circular 
orbital motion was not considered in that investigation.

There has been a growing consensus in recent decades that binarity is 
key for understanding the morphological diversities of pPNe and PNe 
\cite{sok04,sok06}, and considerable efforts have been made to achieve 
a census regarding their companion populations. 
Surveys based on radial velocity variation, photometric variability, 
and infrared excess indicate the presence of companions to the hot 
central stars of PNe with short periods \cite{bon00,mis09,dou15}. 
The detection of long-period binaries relies mainly on direct imaging, 
and thus is difficult as evidenced by the low fraction of known binaries 
in such wide binary systems. 
Detection of companions to the AGB stars is severely hampered due to 
the high obscuration by the thick CSEs, the high brightness contrasts 
of the AGB stars relative to the hypothesized companions, and the 
stellar pulsations of the AGB stars. 
A limited ultraviolet imaging survey has had some success in indicating 
the presence of hot companions or hot accreting material induced by 
the companions \cite{sah08}. 
Long-period AGB binaries are difficult to confirm with the above techniques. 
The spiral-shell patterns propagating through CSEs 
serve as indirect 
probes of the central binaries and therefore remedy the observational 
selection effect. 
The patterns further provide key constraints on the stellar and orbital 
parameters (i.e., stellar masses, orbital period, separation, inclination, 
and eccentricity) \cite{kim13}.

The central puzzle remains in the transition of outflow morphology from 
nearly spherically symmetric CSEs of AGB stars to highly asymmetric PNe. 
Interactions between \emph{close} binaries are hypothesized to be 
responsible for breaking the spherical symmetry by facilitating the 
formation of circumstellar or circumbinary disks and thus opening 
the polar directions for the subsequent vigorous ejection of matter 
observed in many pPNe \cite{mor81}. 
However, the typical dynamical timescales between consecutive arcs 
in the CSEs of the corresponding progenitors are best matched by the 
orbital periods of \emph{wide} binaries \cite{su04,cor04,ram16}.
The coexistence of such geometrically distinct structures therefore 
implies the simultaneous and enigmatic presence of both \emph{wide} 
and \emph{close} binary interactions. 
This conundrum may be solved by invoking binary systems having highly 
eccentric orbits. 
Strong binary interactions at the close periastron passages lead to a 
continuing decrease in the periastron separation, eventually resulting 
in the formation of a disk around the companion star that can launch 
collimated outflows and produce inner bipolar/multipolar structures.

Our results highlight the importance of theoretical investigations of 
binary systems having a wide range of eccentricities and orbital periods 
for explaining a broad variety of observational characteristics seen in 
objects during the AGB-PN transition. 
For example, in the short-period regime, an object like V Hydrae, a carbon 
star that is producing bullet-like mass ejections, is understood through 
an eccentric binary model with an orbital period of 8.5 years \cite{sah16}. 
A very strong gravitational interaction involving an episode of strong 
accretion onto the companion is likely produced during the periastron 
passage, causing the collimated mass ejections. 
On the other hand, AFGL 3068 having a long period of $\sim800$\,years does 
not exhibit such a polar outflow phenomenon. With an extreme eccentricity 
($e=0.8$), which is employed in the model presented in this paper, the 
periastron distance is still larger than 30\,AU. 
The companion does not approach the AGB star sufficiently closely to 
produce the strong interaction needed for the formation of a disk 
powering a collimated outflow. The absence of disk formation at such 
binary separation is consistent with the results of hydrodynamic 
simulations \cite{hua13}. 
Another carbon star, RW Leonis, has an orbital period in the intermediate 
range ($\sim300$ years) \cite{kim13}. Its rather broadened outflow and the 
double spiral feature in the central region are proposed as evidence for 
an eccentric binary \cite{kim15}.

AFGL 3068, exhibiting an almost perfect spiral pattern in the HST image, 
has been considered as a binary system in a circular orbit. 
However, our models for the detailed structures 
clearly revealed by the new ALMA observations of this source lead us, 
for the first time, to the conclusion that a binary system with a highly 
eccentric orbit is responsible for the envelope morphology. This implies 
that binaries in eccentric orbits for stars in the AGB, pPN, and PN phases 
may be ubiquitous over a large period range. Such an expectation follows 
from theoretical considerations as circularization of the orbit due to 
tidal effects is negligible in long-period systems \cite{ver95}.
Starting with this detailed study of AFGL 3068, it is highly desirable 
to determine the binary parameters of a large sample of objects. 
With such a statistical base in hand, one can better assess the role of 
binaries in these important transitional stages of stellar evolution, how 
the envelopes of such binaries are dynamically sculpted, and in particular, 
the role of highly eccentric binary orbits in producing the simultaneous 
presence of different nebular morphologies on a wide range of scales. 

We note that indications of spiral-shell patterns have been observed in the 
circumstellar molecular line images of AGB stars with high sensitivity and 
resolution \cite{cla11,mae12,kim13,ram14,kim15,cer15,dec15}. 
Observations such as these enable detailed investigations 
of the entire nebular structure on spatial scales ranging from several 
to several thousand times the typical size of an AGB star and facilitate 
understandings of the morphological patterns of the AGB, pPN, and PN 
phases. Given early theoretical work on eccentric binary systems revealing 
distortions in the innermost windings of the spiral \cite{rag11}, the high 
angular resolution afforded by ALMA and the Karl G.\ Jansky Very Large Array, 
for example, is ideal for probing the orbital properties of embedded binaries, 
especially their eccentricities. Thus, as higher sensitivity and resolution 
observations of the spatio-kinematic structure through several windings 
of the spiral pattern are obtained for a statistical sample of sources, 
the mechanisms responsible for forming bipolar nebulae may be revealed.

In addition, the circumstellar spiral patterns preserve the fossil 
records of temporal evolution of the mass loss during the AGB phase. 
For R Sculptoris, a giant star with both a spiral and a thermal pulse 
shell, the spiral property was used as a timer for estimating the 
thermal pulsation period \cite{mae12}. For another star, CW Leonis, 
its complex rose-window pattern was understood in terms of a binary 
system but requiring a mass loss variation \cite{cer15}. The full 
intensity images covering all spatial scales will facilitate the 
examination of how binarity affects the return of stellar matter to 
the interstellar medium. Such an interpretation of the circumstellar 
patterns based on binarity and mass loss evolution may apply to other 
systems as well (e.g., the so-called pinwheel structure in Wolf-Rayet 
binaries \cite{tut99}). However, details differ as the shocks in the 
Wolf-Rayet systems are due to colliding winds. 

\clearpage
\begin{figure*} 
  \centering
  \includegraphics[width=\textwidth]{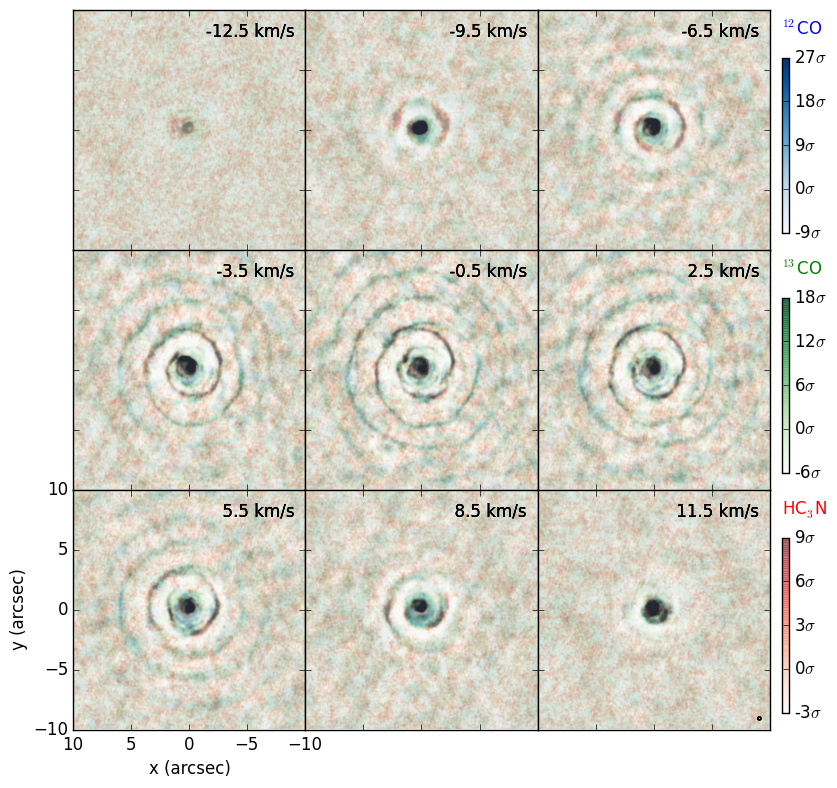}
  \caption{\label{fig:chm}
    {\bf ALMA velocity channel maps of AFGL 3068.} 
    The intensities of three molecular lines $^{12}$CO $J=2-1$ (in blue), 
    $^{13}$CO $J=2-1$ (in green), and HC$_3$N $J=24-23$ (in red) are 
    displayed in all panels, after subtraction of an extended circumstellar 
    component. Each channel is resampled with a spectral width of 3\,\kmps\ 
    and its velocity (relative to the systemic velocity) at the center of the 
    channel is given at the top right side of each panel. The synthesized 
    beam size is denoted at the bottom-right corner of the last panel. North 
    is up, and east is to the left. Color bars indicate the intensity in units 
    of the noise levels for the channel width of 3\,\kmps; $\sigma=2$\,mJy 
    beam$^{-1}$ for $^{12}$CO and $^{13}$CO, and $\sigma=1$\,mJy beam$^{-1}$ 
    for HC$_3$N. 
  }
\end{figure*}

\begin{figure*} 
  \centering
  \includegraphics[width=\textwidth]{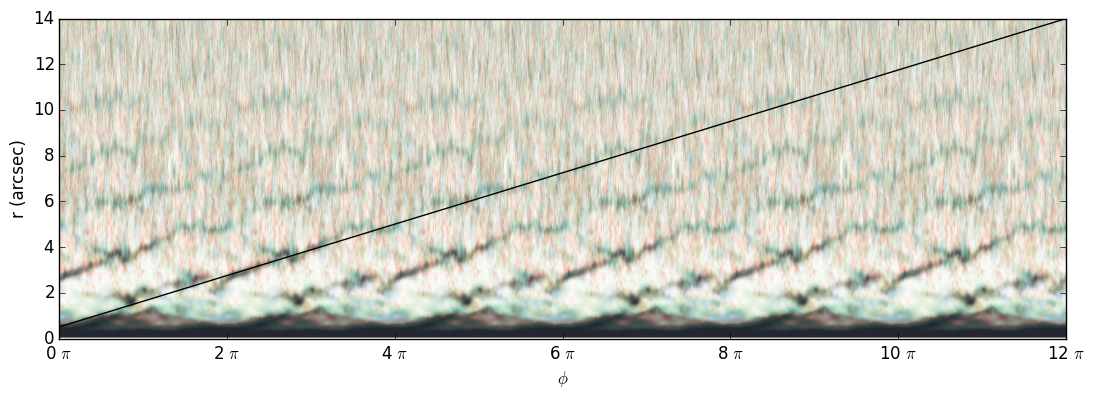}
  \caption{\label{fig:trd}
    {\bf Systemic velocity channel of AFGL 3068 in angle-radius plot.} 
    Same as the middle panel of Fig.\,\ref{fig:chm}, but along the axes 
    of radius from the continuum center versus angle. The angle $\phi$ 
    is measured from the West ($-x$) in the counterclockwise direction. 
    A straight line ($r_{\rm arcsec} = 0.5+9\,\phi/8\pi$) is inserted 
    to guide the eyes to an Archimedean spiral, expected for a binary 
    system in a circular orbit with the orbital plane located exactly 
    on the plane of the sky. A reasonably good match of the observed 
    pattern with this line indicates that an orbital period of 
    $\sim$\,800\,years\,($d$/kpc)\,($V_{\rm wind}$/14\,\kmps)$^{-1}$; 
    its derivation is given in Methods. The deviation of the observed 
    pattern from the straight line is interpreted as deviation of the 
    binary orbit from a circle (see Fig.\,\ref{fig:cmp}).
  } 
\end{figure*}

\begin{figure*} 
  \centering
  \includegraphics[width=\textwidth]{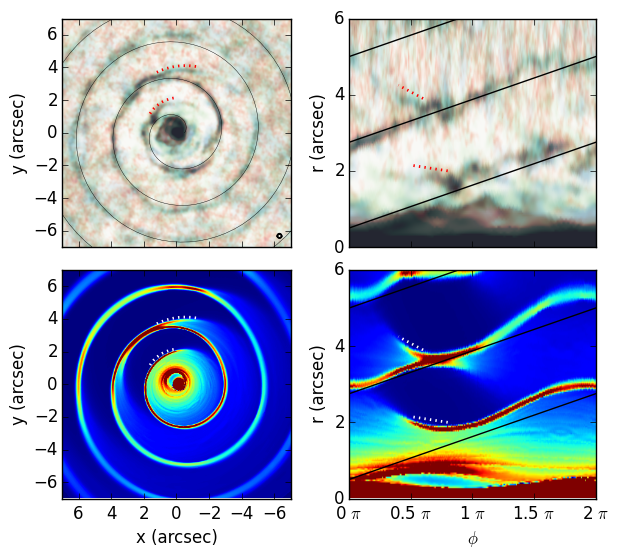}
  \caption{\label{fig:cmp}
    {\bf Bifurcation and undulation of AFGL 3068, and an eccentric binary 
    model.} 
    Comparison between {\bf (a,b)} the observed molecular line intensity 
    at the systemic velocity and {\bf (c,d)} the corresponding gas mass 
    distribution within the same velocity range in a hydrodynamic model 
    employing an eccentric binary with an eccentricity of 0.8 and an 
    inclination angle of 50\arcdeg\ for the orbital plane with respect 
    to the plane of the sky. 
    The straight line is the same as in Fig.\,\ref{fig:trd}. The features 
    related to the bifurcation at two innermost windings of the spiral are 
    marked by dotted lines (in red for the observational data, and in white 
    for the model). The synthesized beam size of the ALMA observations is 
    denoted as an ellipse at the bottom right side of the first panel. 
  } 
\end{figure*}

\clearpage

\section*{Methods}
\subsection*{ALMA observations}

This paper is based on observations taken with the Atacama Large 
Millimeter/Submillimeter Array (ALMA) located in Chile. ALMA is 
currently the most sensitive and flexible millimeter/submillimeter 
interferometer, and it is currently in its Cycle 4 science operation. 
The program under project code 2013.1.00179.S, targeted at AFGL 3068 
(LL Pegasi), was carried out with ALMA on August 29, August 31, and 
September 29 of 2015 during its Cycle 2. The integration time on 
AFGL 3068 was $\sim1.8$ hours, with the total observing time (including 
calibrators and overhead) $\sim3.5$ hours. The 5-field Nyquist-sampled 
mosaic observations used the 12-m array with 34 antennas on average. 
The correlator was configured with four spectral windows (SPWs) 
centered at $\sim$\,230.5, 232.0, 220.4, and 217.8 GHz, which fall in 
the ALMA Band 6 frequency coverage. The bandwidths of two SPWs for the 
$^{12}$CO and $^{13}$CO lines were 117.2\,MHz, while those of the other 
two SPWs were individually set to 1875\,MHz to maximize the sensitivity 
for continuum emission. 
The spectral resolutions of the SPWs including $^{12}$CO and $^{13}$CO 
lines were 488.281\,kHz ($\sim\,0.7\,\kmps$), and those of the other 
two SPWs were 976.563\,kHz ($\sim\,1.3\,\kmps$). 

Manual calibration was performed on the data according to the standard 
procedures of ALMA using the Common Astronomy Software Applications (CASA) 
package \cite{mcm07}. 
For August 29 and 31, J2232+1143, Ceres, and J2253+1608 were used as 
bandpass, flux, and phase calibrators, respectively. For September 29, 
J2253+1608 was used as both bandpass and phase calibrators, and 3c454.3 
was the flux calibrator. 
The use of Ceres as the flux calibrator allowed a high precision flux 
calibration, obtaining the calibrated flux of the phase calibrator
J2253+1608 ($18.45\pm0.04$\,Jy at $\sim\,230.5$\,GHz) with a difference 
between August 29 and 31 less than 1\%. The calibrated flux of J2253+1608 
($14.233\pm0.004$\,Jy) from our data taken on September 29 is consistent 
with the ALMA database value to within 3\% based on calibrator monitoring 
measurements taken on September 28. 

The dataset was further self-calibrated by employing the following procedure 
\cite{pea84}. A continuum map was first made with the line-free channels, 
exhibiting a compact source with a signal to noise ratio greater than 300.
Two cycles of self-calibration against the continuum image were then applied 
to derive new phase and phase-amplitude gains. We achieved an improvement in 
noise level by 20\%, obtaining a final r.m.s.\ noise of 0.2\,mJy per beam in 
the continuum map. The synthesized beam was $0\farcs35\times0\farcs34$ with 
a robust weighting of 0.5.

The resulting calibration gains from self-calibration were applied to 
the line data for further imaging. 
We defined individual cleaning boxes on a channel by channel basis in 
the CLEAN task to properly account for the complex emission region. 
The resulting synthesized beam sizes for all molecular line images 
are about $0\farcs26\times0\farcs23$ with a robust weighting of 0.5.

Supplementary Figs \ref{fig:12CO}--\ref{fig:HC3N} present the channel maps 
and angle--radius plots of $^{12}$CO $J=2-1$, $^{13}$CO $J=2-1$, and HC$_3$N 
$J=24-23$ obtained by the ALMA observations.

\subsection*{Orbital period from an Archimedean spiral in a circular orbit 
approximation}\label{sec:period}

The orbital period of a binary system is simply a function of total 
mass and the summation of semi-major axes of binary orbits (Kepler's 
third law). Therefore, its estimate in a circular orbit approximation 
is valid for eccentric orbit cases. 

The spiral pattern in the CSE of mass-losing star forms at the location of 
mass ejection (i.e., $r=r_{\rm orb}$) with the balance between the radial 
velocity of the wind (i.e., $V_{\rm wind}$) and the tangential velocity 
dragging the wind gusts to the direction of orbit (i.e., $V_{\rm orb}$). 
This statement corresponds to the following equation:
\begin{equation}
  \frac{1}{r_{\rm orb}}\frac{{\rm d}r}{{\rm d}\phi}
  = \frac{V_{\rm wind}}{V_{\rm orb}}
\end{equation}
Assuming a circular orbit and a constant inherent speed of the wind, 
it yields an Archimedean spiral, $r=A+B\,\phi$, where $A$ and $B$ are 
constant values. 
The $A$ parameter indicates the standoff distance of the spiral or 
alternatively provides the position angle of the starting point of 
the spiral, which is in general difficult to identify because of 
complexity in the innermost region of the CSE due to a combination 
of small-scale anisotropy of the mass loss, molecular chemistry and 
insufficient sensitivity. 
The $B$ parameter can be easily measured from the global slope of the 
observed emission pattern in the angle--radius plot, assuming a perfect 
Archimedean spiral. 
For instance, a slope ${\rm d}r/{\rm d}\phi=9\arcsec/8\pi$ characterizes 
the spiral pattern of AFGL 3068 observed by ALMA to first order (as seen 
in Fig.\,\ref{fig:trd}). 
This slope gives a rough estimate for the orbital period of AFGL 3068, 
$T_{\rm orb}\sim$\,800\,years\,($d$/kpc)\,($V_{\rm wind}$/14\,\kmps)$^{-1}$, 
which is derived straightforward from the definitions of orbital period 
$T_{\rm orb}=2\pi r_{\rm orb}/V_{\rm orb}$ and a length unit, parsec (pc), 
$d\rm/pc=arcsec/AU$, where $d$ is the distance to the astronomical object. 

\subsection*{Geometrical model for the bifurcation appearing in an eccentric 
binary}

Supplementary Fig.\,\ref{fig:pw} illustrates how the bifurcation feature 
observed along the circumstellar spiral pattern of AFGL 3068 forms in an 
eccentric binary. The net velocity of the mass-losing AGB stellar envelope 
is the vector sum of the inherent wind velocity (measured in the inertial 
frame of the AGB star) and the orbital velocity of the star. The wind speed 
measured from an observer therefore is fastest in the direction of motion 
of the orbit and slowest in the opposite direction. In a circular orbit 
case, such a wind ejection trend does not change along the orbital passage 
and results in an Archimedean spiral pattern. In an eccentric orbit, however, 
the orbital speed changes along the orbit with its maximum at the periastron. 
To better visualize this effect in Supplementary Fig.\,\ref{fig:pw}, we use 
circular rings to indicate the locations of wind gusts ejected from the AGB 
star at different passages through the orbit revolution. The wind gusts 
ejected at the periastron in the direction of motion plunge through the 
edge of the pattern formed in an earlier orbital cycle. Such overlap of two 
structures appears as a bifurcation, repeating every cycle, and serving as 
distinctive characteristics of eccentric orbits. 

\subsection*{Hydrodynamic model}

The observed spiral pattern in the CSE of a mass-losing star in a given 
eccentric binary model depends on eight key parameters. Among these, five 
parameters (masses of two individual stars, separation, inclination angle, 
and position angle of the current location of the mass-losing star) have 
been investigated in a parameter space analysis for circular binary models 
\cite{kim12b}. An eccentric binary model requires three additional key 
parameters (eccentricity, position angle of the periastron, and position 
angle of the node of the orbit). 

In this paper, we adopt the stellar masses and their mean separation 
derived from a parameter space analysis under a circular orbit assumption 
\cite{kim12b} and introduce an orbital eccentricity as an extra parameter. 
Hence, the binary system that we employ for a hydrodynamic simulation 
consists of a mass-losing star and its companion with the masses of 
3.5\Msun\ and 3.1\Msun, respectively, a mean orbital separation of 
166\,AU, and an eccentricity $e=0.8$. 
The wind is quickly accelerated to reach 14\,\kmps\ well within the 
region of our interest, therefore the modulation in the result is 
certainly due to the orbital motion. 
The hydrodynamic simulation is performed using the FLASH4.3 code with 
an adaptive mesh refinement \cite{fry00} based on a piecewise parabolic 
method \cite{col84}.

To reproduce the projected view of observed pattern, we orient the modeled 
density cube with four angular parameters described above (one inclination 
angle measured from the plane of the sky, and three position angles on the 
plane of the sky) based on alignment of bifurcations, elongation of spiral, 
undulation features in several channels, etc. In particular, the orbit in 
the model displayed in this paper is inclined by 50\arcdeg\ from the plane 
of the sky, producing an additional undulation atop the undulation due to 
the eccentric orbit. The mass-losing star is located at 72\arcdeg\ ahead 
of the periastron in the clockwise direction in the orbital plane. After 
applying the inclination, the density cube is rotated by 130\arcdeg\ in the 
counterclockwise direction. The periastron, set to be on the line of nodes, 
is therefore located along the position angle of 40\arcdeg\ (measured from 
North to the East in the plane of the sky) in Fig.\,\ref{fig:cmp}.

The total stellar mass and mean separation (semi-major axis of the stars) 
would not change significantly with the inclusion of orbital eccentricity, 
because these two parameters are tightly linked to the orbital period, which 
is measured from the arm-to-arm intervals. 
The uncertainty for the mass is larger than that of the semi-major axis 
(Kepler's law). 
The orbital eccentricity and inclination determine the degree of undulation 
in the angle-radius plot, while the three position-angle parameters determine 
the alignment of bifurcation features, the variation of undulation amplitude 
within each winding, the elongation direction of the overall spiral pattern, 
and some details such as the variation of arm-to-arm separation as a function 
of position angle. 

We focus on the unique characteristics of eccentric binaries, revealed by 
the ALMA observation in this paper, from an insightful comparison of the 
pattern produced by a binary in an eccentric orbit with the corresponding 
circular orbit case studied earlier \cite{kim12b}. The bifurcation feature 
becomes distinguishable with an eccentricity $\gtrsim0.5$. The angle between 
the two branches indicates that the eccentricity is significantly larger 
than 0.5 with a good match to the $e\sim0.8$ model. The location of the 
outer branch is mainly determined by the eccentricity, while the location 
of the inner branch highly depends on the inclination angle. The overall 
undulation seen in the angle--radius plot results from the combined effect 
of the eccentricity and inclination angle of the orbit, with a greater 
sensitivity to the eccentricity. We defer a parameter study for a thorough 
modeling to a future investigation.\\

\setlength{\parindent}{0in}

\textbf{Acknowledgments}\\
This paper makes use of the following ALMA data: ADS/JAO.ALMA\#2013.1.00179.S. 
ALMA is a partnership of ESO (representing its member states), NSF (USA) 
and NINS (Japan), together with NRC (Canada), NSC and ASIAA (Taiwan), and 
KASI (Republic of Korea), in cooperation with the Republic of Chile. The 
Joint ALMA Observatory is operated by ESO, AUI/NRAO and NAOJ. 
H.K.\ acknowledges support through East Asian Core Observatories Association 
(EACOA) Fellowship, and thanks F.~Kemper for encouraging the project and 
reviewing an early version of manuscript. 
R.S.'s contribution to the research described here was carried out at the 
Jet Propulsion Laboratory, California Institute of Technology, under a 
contract with NASA, with financial support in part from NASA/STScI HST 
award (GO 11676.02).\\

\textbf{Author contributions}\\
H.K.\ planned the project, prepared and submitted the proposal, and wrote 
the manuscript. 
A.T.\ was involved in observation preparation, data reduction and analysis, 
and commented on the manuscript. 
S.-Y.L.\ was involved in project planning, data interpretation, and manuscript 
preparation.
R.S., R.E.T., M.M.\ and N.H.\ were involved in the science discussion as well 
as writing the proposal and manuscript. 
I.-T.H.\ did the radiative transfer modeling in the proposal preparation that 
generated the data for this study.\\

\textbf{Data availability}\\
The data that support the plots within this paper and other findings of this 
study are available from the corresponding author upon reasonable request.\\

\textbf{Additional information}\\
Correspondence and requests for materials should be addressed to Hyosun Kim 
(hkim@asiaa.sinica.edu.tw).

\setcounter{figure}{0}
\renewcommand{\figurename}{{\bf Supplementary Figure}}

\begin{figure*} 
  \centering
  \includegraphics[width=0.96\textwidth]{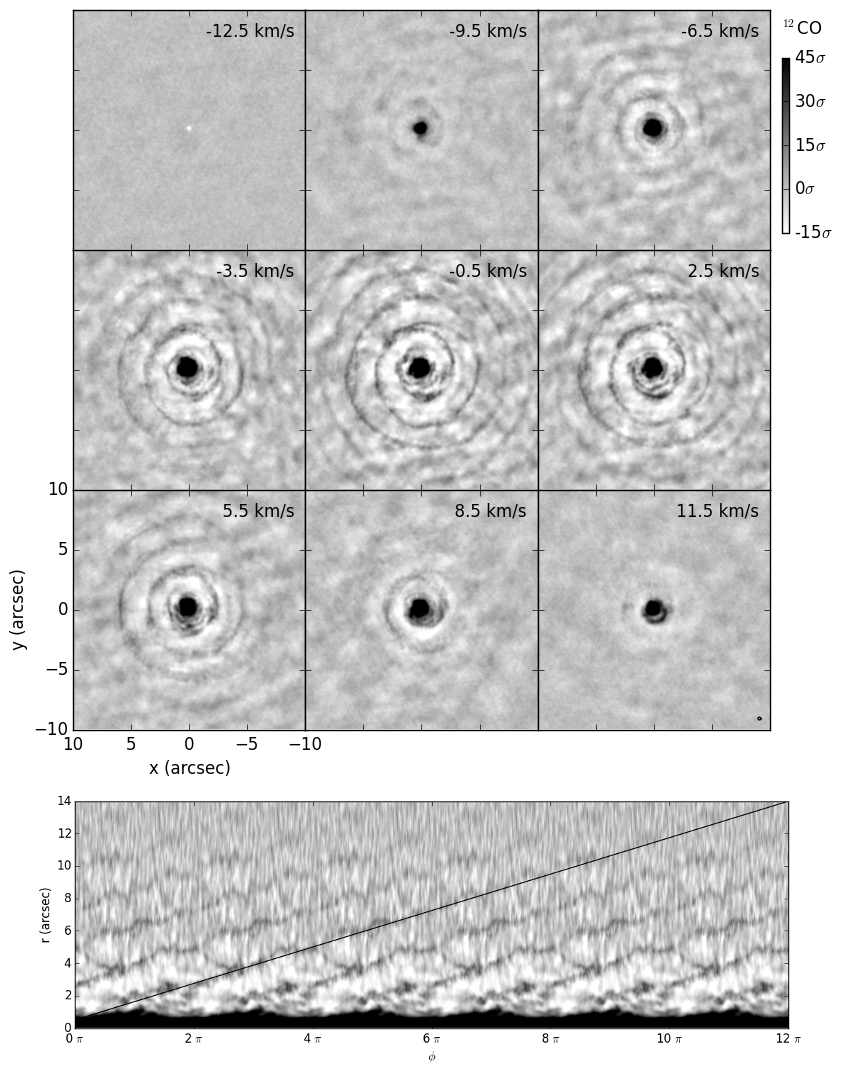}
  \caption{\label{fig:12CO} 
    {\bf\boldmath $^{12}$CO $J=2-1$ of AFGL 3068.} The channel map (top) 
    and angle--radius plot (bottom) of $^{12}$CO molecular line emission. 
    See Figs \ref{fig:chm} and \ref{fig:trd} captions for details.}
\end{figure*}

\begin{figure*} 
  \centering
  \includegraphics[width=0.96\textwidth]{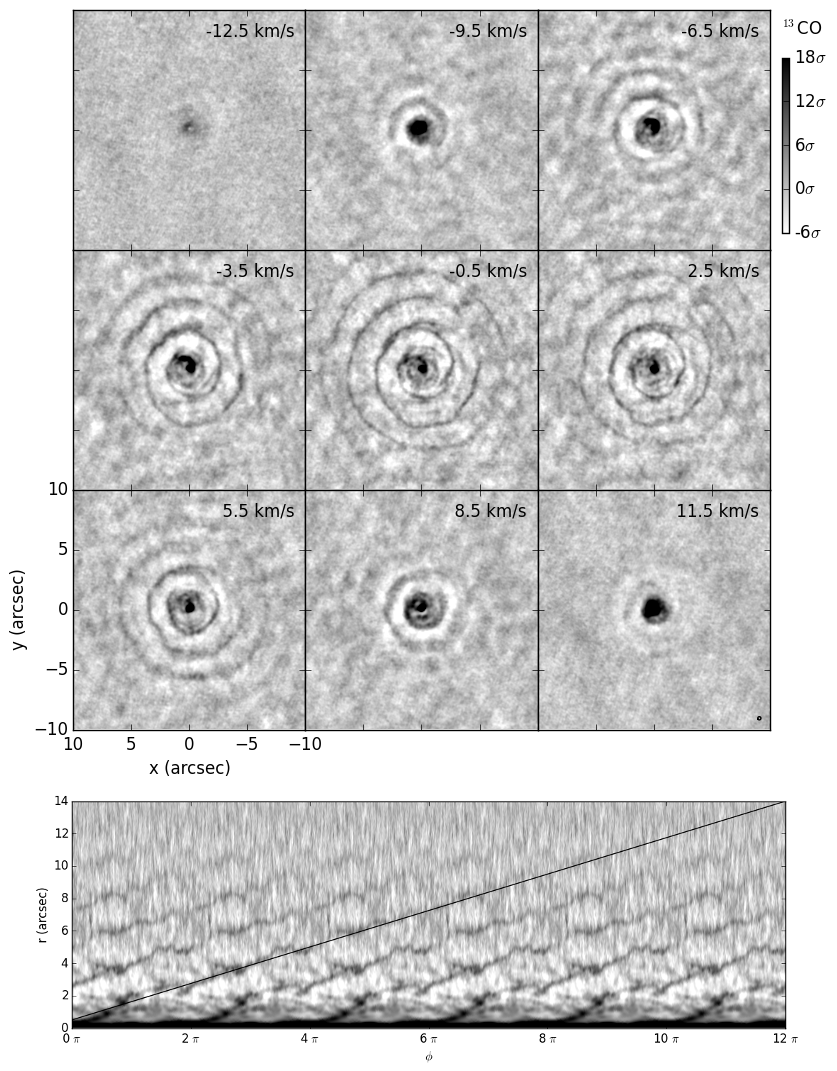}
  \caption{\label{fig:13CO} 
    {\bf\boldmath $^{13}$CO $J=2-1$ of AFGL 3068.} The channel map (top) 
    and angle--radius plot (bottom) of $^{13}$CO molecular line emission. 
    See Figs \ref{fig:chm} and \ref{fig:trd} captions for details.}
\end{figure*}

\begin{figure*} 
  \centering
  \includegraphics[width=0.96\textwidth]{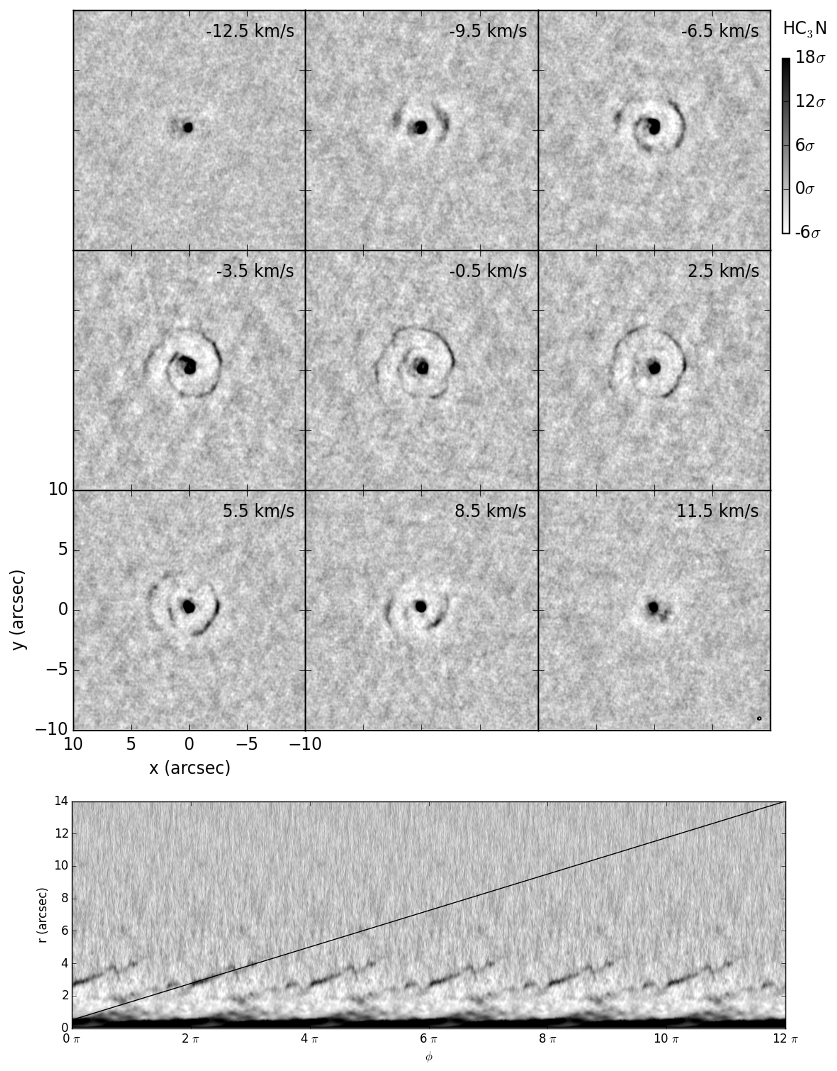}
  \caption{\label{fig:HC3N} 
    {\bf\boldmath HC$_3$N $J=24-23$ of AFGL 3068.} The channel map (top) 
    and angle--radius plot (bottom) of HC$_3$N molecular line emission. 
    See Figs \ref{fig:chm} and \ref{fig:trd} captions for details.}
\end{figure*}

\begin{figure*} 
  \centering
  \includegraphics[width=\textwidth]{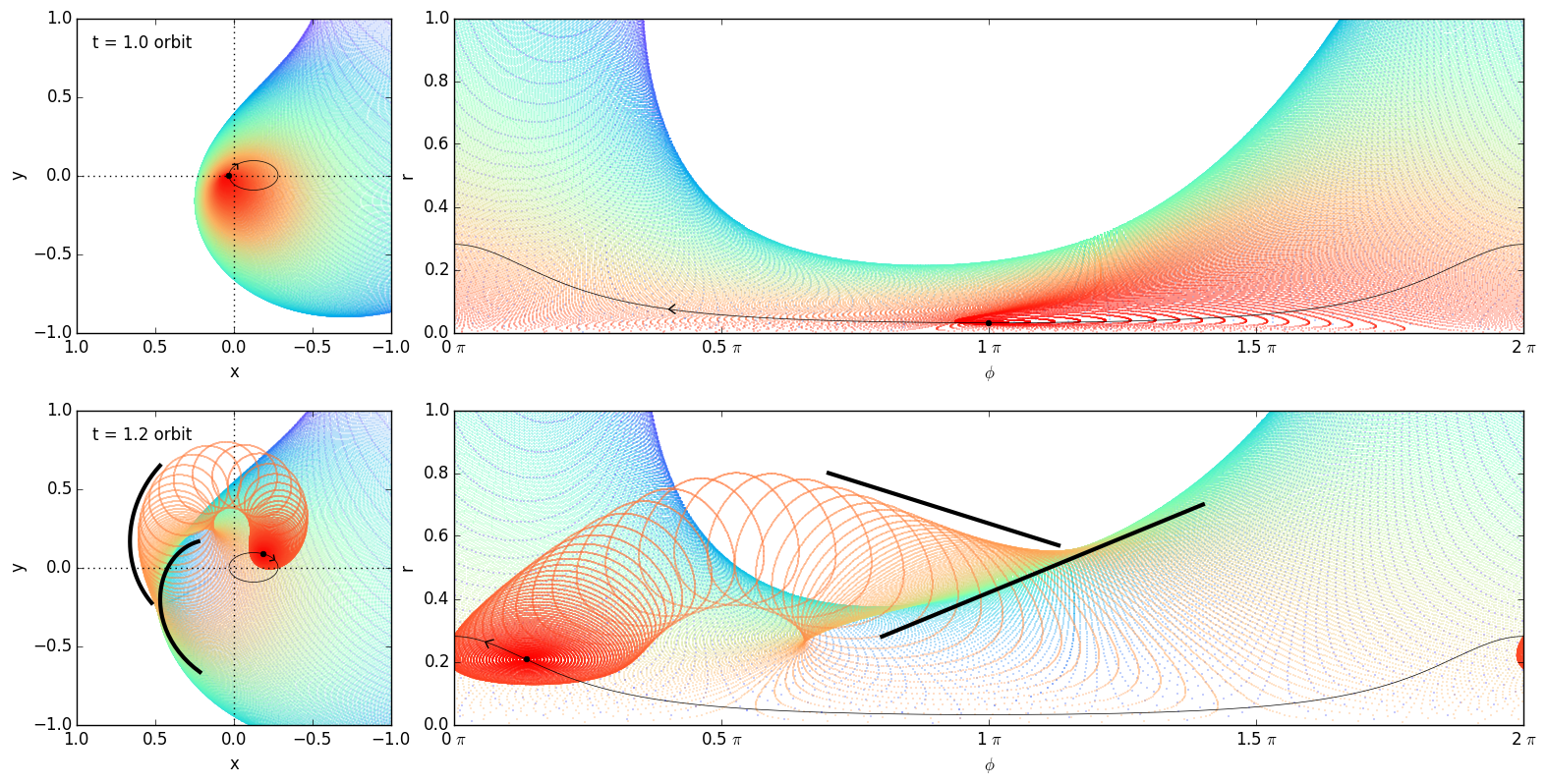}
  \caption{\label{fig:pw} 
    {\bf\boldmath Formation of a bifurcation in an eccentric binary.}
    A geometrical model illustrating the distributions of wind material 
    from a mass-losing star in the $x$--$y$ plane (left) and $\phi$--$r$ 
    plane (right) after 1 orbit (top) and 1.2 orbits (bottom) starting 
    from the periastron of its orbit in the clockwise direction (marked by 
    arrows). The origin of coordinates at left is at a focus of the ellipse 
    (thin line) representing the relative location of the mass-losing star 
    (eccentricity of 0.8). The faster wind in the forward direction of 
    motion at the periastron overtakes the material distributed during the 
    earlier orbit, displaying the bifurcation. Thick lines trace the feature 
    of interest. Red to blue colors represent the ejection time sequence of 
    wind gusts from recent to the past.}
\end{figure*}

\setcounter{figure}{0}
\renewcommand{\figurename}{{\bf Supplementary Video}}

\begin{figure*} 
  \centering
  \caption{
    {\bf\boldmath Visualizing the ALMA image cube of AFGL 3068.} Each frame 
    of the video shows the composite image of $^{12}$CO $J=2-1$, $^{13}$CO 
    $J=2-1$ and HC$_3$N $J=24-23$ emission for a different line-of-sight 
    velocity. This velocity, relative to the systemic velocity, advances by 
    1\,\kmps\ per frame, and is given at the top-right corner. The presented 
    field area is $10\arcsec\times10\arcsec$. North is up and east is to the 
    left. (Video available from Nature Astronomy website.)}
\end{figure*}

\end{document}